\documentclass[aps,apl,twocolumn,superscriptaddress,showpacs]{revtex4-1}
\usepackage{latexsym}
\usepackage{graphicx}
\usepackage{amsmath}
\usepackage{natbib}
\begin{document}

\title{Silicon Superconducting Quantum Interference Device}

\author{J. E. Duvauchelle}
\affiliation{Univ. Grenoble Alpes, CEA - INAC - SPSMS, F-38000 Grenoble, France}
\author{A. Francheteau}
\affiliation{Univ. Grenoble Alpes, CEA - INAC - SPSMS, F-38000 Grenoble, France}
\author{C. Marcenat}
\affiliation{Univ. Grenoble Alpes, CEA - INAC - SPSMS, F-38000 Grenoble, France}
\author{F. Chiodi}
\affiliation{Univ. Paris-sud, CNRS - IEF, F-91405 Orsay - France}
\author{D. D\'ebarre}
\affiliation{Univ. Paris-sud, CNRS - IEF, F-91405 Orsay - France}
\author{K. Hasselbach}
\affiliation{Univ. Grenoble Alpes, CNRS - Inst. N\'eel, F-38000 Grenoble, France}
\author{J. R. Kirtley}
\affiliation{Center for probing at nanoscale, Stanford University, Palo Alto, California 94305-4045, USA}
\author{F. Lefloch}
\affiliation{Univ. Grenoble Alpes, CEA - INAC - SPSMS, F-38000 Grenoble, France}
\email[Corresponding author: ]{francois.lefloch@cea.fr}

\begin{abstract}
We have studied a Superconducting Quantum Interference SQUID device made from a single layer thin film of superconducting silicon. The superconducting layer is obtained by heavily doping a  silicon wafer with boron atoms using the Gas Immersion Laser Doping (GILD) technique. The SQUID device is composed of two nano-bridges (Dayem bridges) in a loop and shows magnetic flux modulation at low temperature and low magnetic field. The overall behavior shows very good agreement with numerical simulations based on the Ginzburg-Landau equations.        
\end{abstract}
\maketitle

Superconducting devices are currently used in many applications where they have shown their great potential. This ranges from detectors for astrophysics up to a single photon level detection \cite{Natarajan2012} to very sensitive magnetic sensors with Superconducting QUantum Interference Devices SQUIDs \cite{Fagaly2006}.  Another promising and rapidly growing field of research where superconducting devices play an important role is quantum information. So far, the most advanced circuits for quantum computing are based on superconducting qubits \cite{Devoret2013}. Many efforts have been made to increase the number of qubit operations that can be made within the quantum coherence time, including improvements using quantum error correction schemes and quantum non destructive measurements \cite{Devoret2013}. Even though multiple coupled qubit manipulations have already been  performed \cite{Reed2012, Kelly2015}, the next step is to demonstrate the possibility of performing operations that involve multiple logical qubits. Another candidate for quantum information is the spin qubit \cite{Loss1998}. First realizations were obtained with two dimensional electron gas \cite{Petta2005, Koppens2006} and recently quantum dots made from isotopically purified silicon have shown very long coherence times \cite{Saeedi2013, Muhonen2014}. All this future developments will require more and more complex architectures for which the technology must be perfectly controlled in terms of reproducibility, variability and know-how. In addition, the quality of the underlying material must also be addressed as it can be an important source of decoherence for quantum devices \cite{Martinis2015}. In that quest, superconducting boron doped silicon discovered in 2006 \cite{Bustarret2006}, can address all of these challenges at once as silicon technology is by far the most mature one for nanoelectronics and for which extremely high quality material can be used \cite{Shim2014}. Moreover, the relatively high normal state resistivity of boron doped silicon allows using this material for Microwave Kinetic Inductance Detection \cite{Shim2015}. 

As a first step towards superconducting silicon based quantum engineering, we have fabricated a SQUID (Superconducting Quantum Interference Device) from a single layer of superconducting silicon.  Our results show a flux modulation at low temperature and low field that demonstrates the macroscopic nature of the quantum wave function of the superconductivity in silicon \cite{Likharev1979,Tinkham1996,Barone1982, Orlando1991}. Our results confirm that boron doped silicon can be exploited for applied superconducting devices and quantum technology.

The SQUIDs are composed of two weak links (Dayem bridges) acting as two Josephson junctions \cite{Anderson1964, Hasselbach2000} and are processed from a single layer of superconducting silicon. The superconducting film has been obtained by heavily doping the silicon with boron atoms using the Gas Immersion Laser Doping technique \cite{Kerrien2002}. This technique consists of shining laser pulses (here an XeCl $308 \,nm$ laser with a $25 \,ns$ pulse duration) at the surface of a silicon wafer on top of which molecules of $BCl_3$ have been previously chemisorbed. During a laser pulse, the silicon melts over a thickness and a time that depend on the energy density of the laser. In this melting time, the boron atoms diffuse very rapidly into the melted phase while the chloride atoms are expelled. At the end of the laser pulse, the boron atoms are incorporated in substitutional sites as the Si:B layer is epitaxially grown over silicon. Since the number of atoms that can be chemisorbed at the surface of silicon is self-limiting, the number of boron atoms introduced at each laser pulse is very reproducible and equals $1.2\times10^{14}$ cm$^{-2}$. In order to increase the amount of dopants, the overall procedure (gas immersion and laser shot) is repeated (typically 10 to few hundred times) to reach active dopant concentrations up to $11 \%\,at$ with no boron aggregates \cite{Kerrien2002, Dubois2008, Bhaduri2012}. Such a high level of concentration, larger than the solubility limit of boron in silicon ($\simeq 1 \%\,at$), can be achieved thanks to the rapid liquid/solid phase transition that quenches the boron atoms into the crystalline phase. 
In order to control the melting depth, the melting duration is monitored through the time resolved silicon reflectivity using a low power laser at 675 nm during each laser pulse. This procedure produces a very sharp doped-undoped interface of only a few $nm$ \cite{Hoummada2012}. In our set-up, the surface area covered by the laser beam is $2 \,mm \times 2 \, mm$. The study of the electronic transport properties down to very low temperature has revealed superconducting properties for a boron dose larger than $4\,\times 10^{15} \,cm^{-2}$ with a maximum of the critical temperature below which Si:B thin films show a zero resistance state of $0.7 \,K$ for a boron concentration of $10\, \%\, at$ and a thickness of $200 \,nm$ \cite{Marcenat2010, Grockowiak2012, Chiodi2014}.    

The superconducting silicon layer we have used for the present study was grown with 200 laser pulses and a melting time duration monitored to $47 \,ns$. With these conditions, we obtained a $80 \,nm$ thick layer of boron doped silicon Si:B with a boron dose of $2.4\,\times10^{16}cm^{-2}$, which corresponds to a concentration of $5 \%\,at$. 

\begin{figure}[ht]
\includegraphics[width=0.4\textwidth]{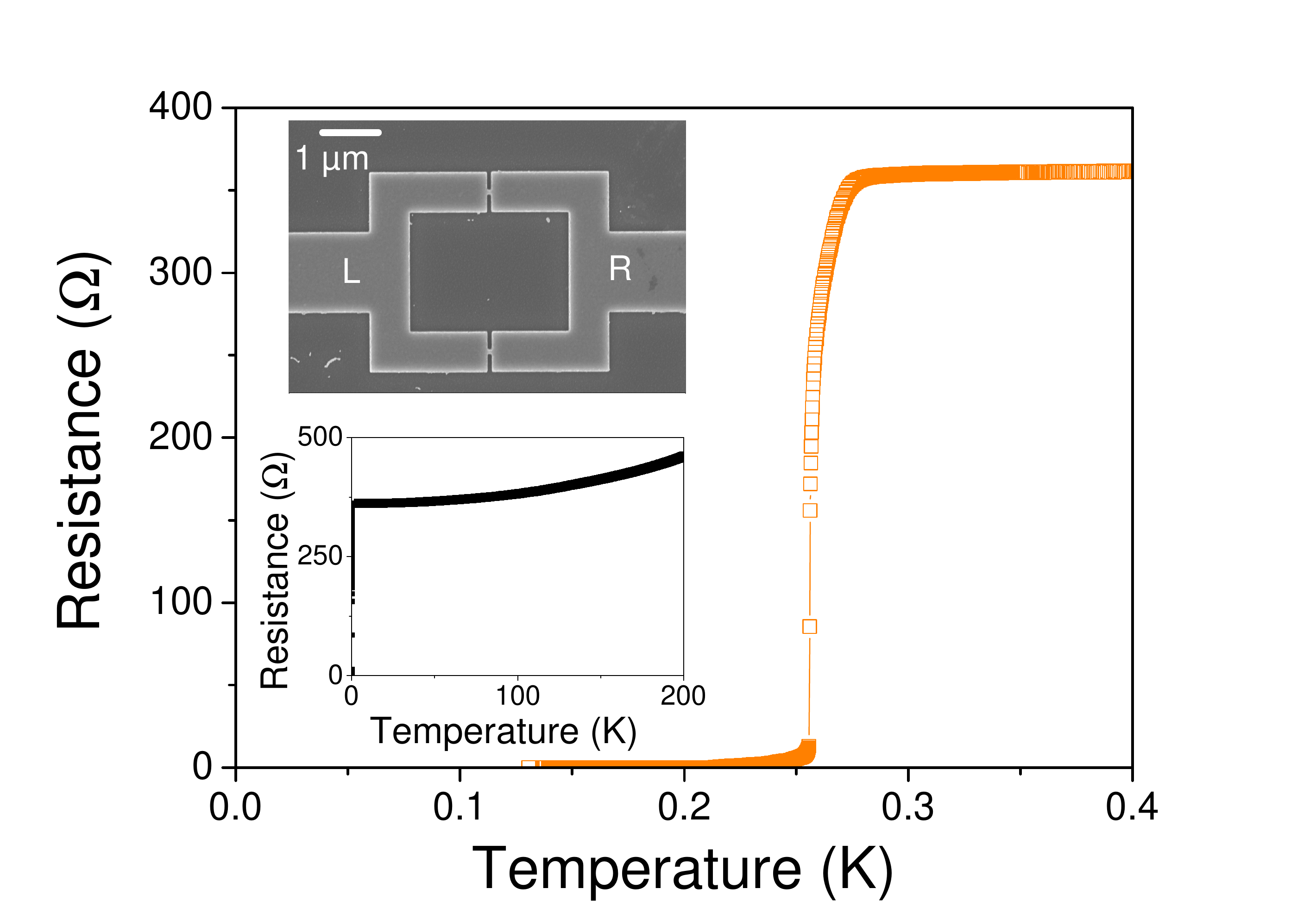}
\caption{Resistive transition of a superconducting silicon SQUID. The SEM image shows the actual SQUID fabricated from a bare superconducting Si:B doped layer. It is composed of two weak links (Dayem bridge) with dimensions of 100 nm x 100 nm for the actual device reported here (see left image). Insert : Resistance of the device in the high temperature range showing a clear metallic behavior.}
\label{Fig1}
\end{figure}

The fabrication of the SQUIDs starts with electron beam lithography to define the shapes (loops, weak links and contact electrodes) of the devices. Then, an aluminum hard mask (20 nm) is deposited by lift-off. More than $100 \, nm$ of the silicon not protected by the aluminum layer is etched away by reactive ion etching using fluoride gas and a highly anisotropic recipe. The aluminum mask is entirely removed afterwards by wet etching before the deposition of Ti/Au contact pads defined by optical lithography. Care has been taken to deoxydize the doped silicon before the deposition of the contact pads. A SEM image of a typical realization is shown in Fig. \ref{Fig1} (left insert). The SQUIDs all have the same loop area and the dimensions of the weak links range from $80 \,nm$ to $200 \,nm$ in width and from $100 \,nm$ to $500\,nm$ in length. In this letter we present the results obtained in a SQUID with weak links of $100 \, nm$ by $100 \, nm$, but all the devices we have measured show magnetic flux modulation. We have also checked that the flux modulation in a SQUID loop can be reproduced using different silicon batches. 

Fig.\ref{Fig1} shows the resistance of a SQUID as a function of temperature. The transition to the superconducting zero resistance state occurs at $T_c=260 \,mK$ as expected for such a Si:B layer. The right insert of Fig.\ref{Fig1} shows the resistance of the same device at high temperature up to $250 \,K$. The behavior is clearly metallic with a residual resistive ratio $R_{300\,K}/R_{10\,K}\simeq 1.4$.   

\begin{figure}[ht]
\includegraphics[width=0.4\textwidth]{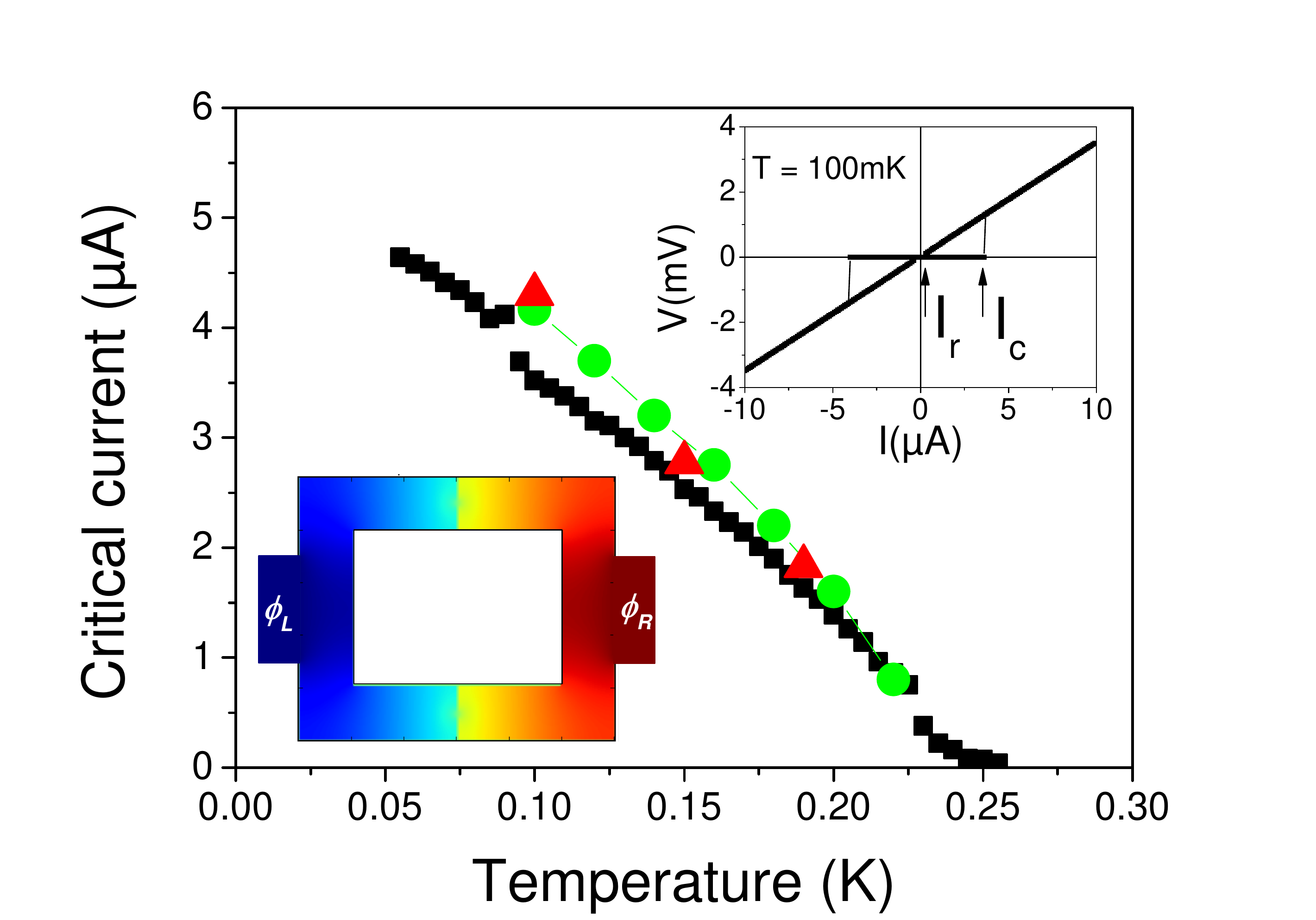}
\caption{Temperature dependence of the measured critical current of the silicon SQUID (black square points). The green circles correspond to the maximum critical current obtained by measuring the magnetic flux modulation and the red triangles result from solving the Ginzburg-Landau equations. Left insert : color plot of the phase evolution of the superconducting wave function along the device for a phase coherence length of $40 \, nm$ and an applied phase fixed to $\pm 3\pi/4$ at both ends of the device. Right insert : IV characteristics at $100 \,mK$ showing a strong hysteresis when ramping the applied DC current. $I_c$ and $I_r$ show the critical current and the re-trapping current respectively.}  
\label{Fig2}
\end{figure}

At low temperature, a SQUID critical current $I_c$ of several $\mu A$ (corresponding to a current density of $\approx 10^4 \,A/cm^2$) is observed in the current-voltage I-V characteristics. Below $220 \, mK$ the IV response is strongly hysteretic (see right insert of Fig.\ref{Fig2}) with a very small re-trapping current $I_r$. For current larger than the critical current, the resistance, defined as the slope of the I-V curve, is $360 \,\Omega$, which is exactly that of the device above the critical temperature. This means that the entire structure turns normal in the dissipative state. This behavior is due to a heating effect that propagates from the weak links to the rest of the device, as is usually the case for such devices. The temperature dependence of the critical current measured with no applied field is plotted in Fig.\ref{Fig2}. It shows a rather smooth variation with a critical current vanishing at the critical temperature. 

For superconducting Josephson weak links, the amplitude of the critical current strongly depends on the length of the bridge compared to the superconducting Ginzburg-Landau (GL) coherence length  $\xi$. In the dirty limit $\xi=\sqrt{\xi_0 l}$, where $\xi_0$ is the BCS (Bardeen-Cooper-Schrieffer) coherence length and $l$ the elastic mean free path \cite{Likharev1979,Tinkham1996}. For superconducting silicon at our doping level $l\approx 2-3 \,nm$ and $\xi_0\approx 1000\,nm$, which gives the zero temperature phase coherence length $\xi(0)\approx 40-50 \,nm$ \cite{Marcenat2010}. In the actual geometry, the ratio between the length of the weak link and the GL coherence length is $L/\xi(0)\simeq 2.5$. 
The critical current of Josephson bridge junctions can be obtained by solving the Ginzburg-Landau equations. Following the code developed by Hasselbach et al. \cite{Hasselbach2000}, we have numerically solved these equations in two dimensions taking into account the exact geometry of our device. This calculation is done by assuming a macroscopic superconducting wave function that can be written as $\psi(r) = f(r)\psi_\infty e^{i\phi(r)}$ where both $f$ and $\phi$ depend on position but $\psi_\infty$ is a constant at a given temperature. The values of $f(r)$ are fixed to $f=1$ at the left and right sides of the devices (position R and L in the SEM image) which means that the amplitude of the order parameter is constant and equals to its equilibrium value in the electrodes. The simulation is done with a certain phase $\phi_R$ and $\phi_L$ imposed at the left and right electrodes (usually $\phi_R=-\phi_L$). We then obtain the amplitude and phase everywhere along the device. For the simulation, we have assumed a perfectly symmetrical case where the critical current in the upper and lower branches are equal. The left insert of Fig. \ref{Fig2} shows a color plot of the local superconducting phase for a total phase difference $\Delta \phi =\phi_L-\phi_R \simeq 3\pi/2$. The simulation shows that the main phase drop occurs at the junction bridges. Doing the same simulation for various phase differences, gives the current-phase relationship of the device and in turn, the magnetic field dependence of the critical current of the SQUID. The temperature dependence is obtained by taking into account the temperature dependence of the GL phase coherence length that modifies the ratio $L/\xi(T)$ with $\xi(T)=\frac{\xi(0)}{\sqrt{1-T/T_c}}$. The results for the critical current at three distinct temperatures is plotted in Fig.\ref{Fig2}. They must be compared, not to the critical current obtained at no applied magnetic field, but to the maximum of the critical current flux modulation at different temperatures. Indeed, due to remanent and earth field contributions, zero applied field does not necessarily mean zero magnetic field. The agreement with experimental data is very good (see Fig.\ref{Fig2}). The general temperature dependence obtained here shows a behavior very close to what has been obtained previously for similar values of $L/\xi(0)$  \cite{Hasselbach2002,Podd2007,Hazra2014}.   

One can also estimate the product $R_N I_{c,wl}$ where $R_N$ is the normal state resistance of one bridge junction and $I_{c,wl}$ its critical current. The square resistance of the doped silicon film at low temperature can be extracted from the normal state resistance of the entire structure taking into account the geometry of the SQUID. Since the bridge is a square, the normal state resistance of one junction is therefore $R_N=R_\Box \simeq 14\,\Omega$, in agreement with the known resistivity of $\rho \simeq 100 \,\mu \Omega.cm$ at such a doping level and thickness. Considering that the SQUID is symmetric, we find $R_N I_{c,wl} \simeq  35\mu V$ with $I_{c,wl}=I_c/2=2.5 \,\mu A$. This value can be compared to the superconducting gap $\Delta / e \simeq 40 \,\mu V$ that we estimate from the critical temperature $T_c=0.26 \,K$ using the BCS relationship $\Delta =1.76 k_B T_c$ \cite{Dahlem2010}. We then obtain $R_N I_{c,wl}\simeq \Delta/e$ in agreement with existing calculations \cite{Likharev1979,Tinkham1996}.

\begin{figure}[ht]
\includegraphics[width=0.4\textwidth]{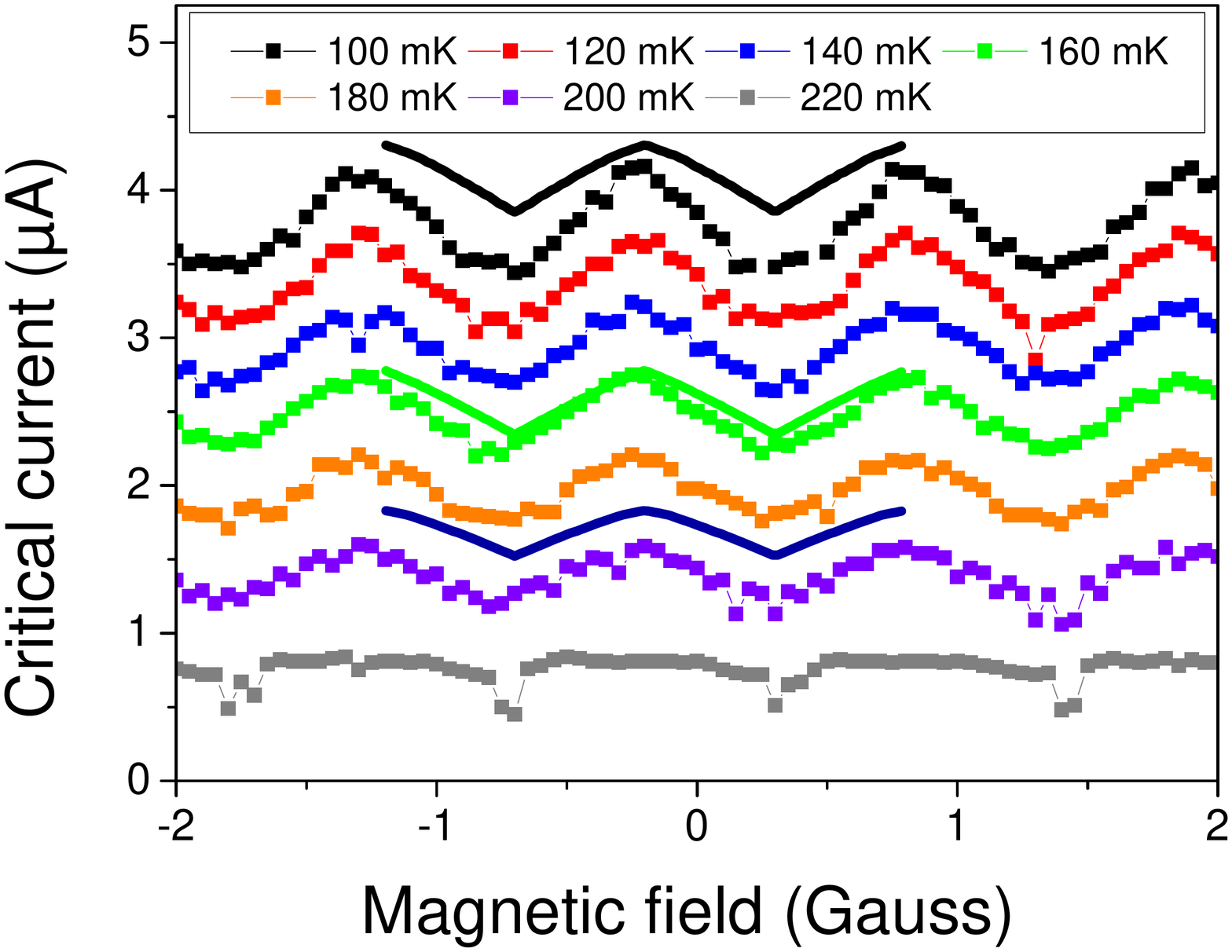}
\caption{Magnetic field dependence of the critical current of the DC SQUID at various temperatures from $100 \,mK$ (top) to $220\,mK$ (bottom). The measured modulation (scattered points) is compared to the flux dependence obtained from the GL equations in the actual geometry for three different values of the phase coherence lengths $40 \,nm, 60 \,nm$ and $80 \,nm$ to account for the temperature dependence ($T\simeq 100 \,mK -\text{top}, 150\,mK -\text{middle and } 190 \,mK - \text{bottom}$) (solid lines).}
\label{Fig3}
\end{figure} 

In Fig.\ref{Fig3}, we show the dependence of the critical current with the magnetic field applied perpendicular to the device, for various temperatures. Each of these curves has been obtained by ramping the magnetic field from $-2 \,G$ to $+2\,G$ in steps of $0.1\,G$. The results show a very regular oscillation of the critical current as a function of the magnetic field with a period of $1\,G$. This value corresponds to a one flux quantum $\Phi_0=h/2e=2\,\times 10^{-15}\,Wb$ in a surface area of $20 \, \mu m^2$, which is the surface of the SQUID loop $4 \,\mu m \times 5\, \mu m$. The amplitude is roughly $10 \%$ of the maximum critical current. This flux (or field) modulation decreases when increasing the temperature and vanishes to zero when approaching the critical temperature. The solid lines in Fig.\ref{Fig3} are the results of the critical current modulation obtained by numerically solving the GL equations. This has been done for three different values of the GL coherence lengths : $40 \,nm, 60\,nm$ and $80 \,nm$ corresponding to $100 \,mK, 150 \,mK$ and $190 \,mK$ approximatively. The position of the curve follows rather well the temperature dependence (as already seen in Fig.\ref{Fig2}) and the amplitude of the modulation is well reproduced too. 

In conclusion, we have measured the magnetic flux dependence of a DC SQUID device made from a superconducting boron doped silicon thin layer. The SQUID geometry consists of a $20 \,\mu m^2$ loop interrupted by two nano-bridges forming two Josephson junctions. We observed clear oscillations of the critical current as a function of the applied magnetic flux. The amplitude of the magnetic flux modulation and the temperature dependence of the critical current are well reproduced by numerically solving the Ginzburg-Landau equations.  These findings are reproducible for samples with various nano-bridge dimensions and fabricated from different Si:B layers. These results demonstrate that superconductivity in silicon can be described by a macroscopic wave function and open routes towards silicon based superconducting quantum engineering. The future will be to implement more practical devices \cite{Shim2014,Shim2015} such as superconducting resonators and JoFET's (Josephson Field Effect Transistors) in which a Josephson current flowing between superconducting drain and source contacts through an undoped channel will be modulated by an electrostatic gate. Using ultimate technology to reduce the size of the channel to only few nanometers, one can expect to obtained tunnel coupling between the drain and the source.          
 
We acknowledge the Grenoble Nanoscience Foundation for the Chair of Excellence of J. R. Kirtley. Samples were fabricated using the IEF and PTA Renatech facilities. The financial support was partially obtained from the French National Research Agency (ANR-NanoQuartet Grant No. NR12BS1000701) and the CEA nanoscience transverse program.


\end{document}